# Local Electronic Structure Changes in Polycrystalline CdTe with $CdCl_2$ Treatment and Air Exposure

*Morgann Berg,[†,\*] Jason M. Kephart,[‡] Amit Munshi,[‡] Walajabad S. Sampath,[‡] Taisuke Ohta,[†] Calvin Chan[†,\*]*

[†] Sandia National Laboratories, Albuquerque, New Mexico 87185, United States

[‡] Department of Mechanical Engineering, Colorado State University, Fort Collins, Colorado 80523, United States

ABSTRACT: Post-deposition $CdCl_2$ treatment of polycrystalline CdTe is known to increase photovoltaic efficiency. However, the precise chemical, structural, and electronic changes that underpin this improvement are still debated. In this study, spectroscopic photoemission electron microscopy was used to spatially map the vacuum level and ionization energy of CdTe films, enabling the identification of electronic structure variations between grains and grain boundaries.

*In vacuo* preparation and inert transfer of oxide-free CdTe surfaces isolated the separate effects of CdCl$_2$ treatment and ambient oxygen exposure. Qualitatively, grain boundaries displayed lower work function and downward band bending relative to grain interiors, but only after air exposure of CdCl$_2$-treated CdTe. Analysis of numerous space charge regions at grain boundaries (GBs) showed an average depletion width of 290 nm and an average band bending magnitude of 70 meV, corresponding to a GB trap density of $10^{11}$ cm$^{-2}$ and a net carrier density of $10^{15}$ cm$^{-3}$. These results suggest that both CdCl$_2$ treatment and oxygen exposure may be independently tuned to enhance CdTe photovoltaic performance by engineering the interface and bulk electronic structure.

INTRODUCTION: Manufacturing efficient cadmium telluride (CdTe) solar cells with > 21% power conversion efficiency requires post-deposition treatment with cadmium chloride (CdCl$_2$).[1] This treatment results in recrystallization of the CdTe layer, producing larger, more uniform grains.[1-4] However, efficiency improvements cannot be attributed solely to microstructural changes because single crystal CdTe exhibits poor photovoltaic performance,[5-7] and CdTe films deposited at high temperature show increased efficiency with CdCl$_2$ treatment despite minimal changes in grain size.[8,9]

Electronic structure changes at grain boundaries (GBs) have also been proposed to result from CdCl$_2$ treatment based on observations of thermally-assisted tunneling,[10] enhanced GB current collection,[11] and electric potential variation at GBs.[5,12,13] The effect of CdCl$_2$ on GBs was described as an electrostatic rigid shift in the electronic energy band positions due to local doping, i.e., the vacuum level ($E_{vac}$), conduction bands ($E_{CB}$), and valence bands ($E_{VB}$) were assumed to shift equally at GBs.

Thus far, however, only the $E_{vac}$ variation at GBs has been determined from surface potential measurements. Without direct measurement of $E_{CB}$ and $E_{VB}$, the band alignment at the GBs remains unknown. Similarly, the effects of dopants and impurities (e.g., sulfur and oxygen[14,15]), introduced during $CdCl_2$ treatment, have not been strictly separated from that of $CdCl_2$. Oxygen incorporation is an implicit step in $CdCl_2$ processing of CdTe devices, introduced during rinsing and annealing of $CdCl_2$-treated CdTe in air.

Here, spectroscopic photoemission electron microscopy (PEEM) was used to probe the effects of $CdCl_2$ treatment and air exposure on the local $E_{vac}$ and ionization energy ($IE = E_{vac} - E_{VB}$) of CdTe films at the nanometer length scale. Preparation of oxide-free CdTe surfaces, inert environment sample transfer, and PEEM measurement in ultrahigh vacuum (UHV) allowed the examination of local electronic properties of untreated and $CdCl_2$-treated CdTe, with and without the influence of oxygen (via air exposure). $CdCl_2$ processing lowered the average $IE$ of CdTe, while air exposure increased the average $IE$. A decrease in work function at GBs relative to grain interiors was observed only after exposing the $CdCl_2$-treated CdTe surface to air. By analyzing numerous space charge regions formed at GBs after air exposure, estimates of the net carrier density and GB trap density were determined. Unlike previously reported studies of CdTe GBs, PEEM provided a direct comparison of the local $IE$ and $E_{vac}$ for $CdCl_2$-treated CdTe.

RESULTS AND DISCUSSION: Figure 1 illustrates the examined superstrate CdTe films, consisting of a glass superstrate coated with a transparent conducting oxide (TCO), 100 nm $Mg_{0.23}Zn_{0.77}O$ (MZO), and 5 μm CdTe.[16] Prior to PEEM imaging, samples were mechanically polished to 3 μm in thickness and cleaned with low-energy ion sputtering to provide smooth, oxide-free surfaces.[17]

X-ray photoelectron spectroscopy showed less than 1 at% oxygen and 17 at% carbon after sputtering.[18] After sputter-cleaning, samples were transferred in N$_2$ (<0.1 ppm O$_2$, <0.1 ppm H$_2$O) to the UHV PEEM measurement chamber, without additional exposure to air. After initial PEEM measurement, samples were exposed to air for 30 minutes, reintroduced into UHV, and examined to investigate the impact of air exposure. Figure 1 also illustrates grains and GBs in the polished surface, and defines the variation between one grain and another (grain-to-grain), and between a grain and GB (grain-to-boundary), referred to in the following discussion.

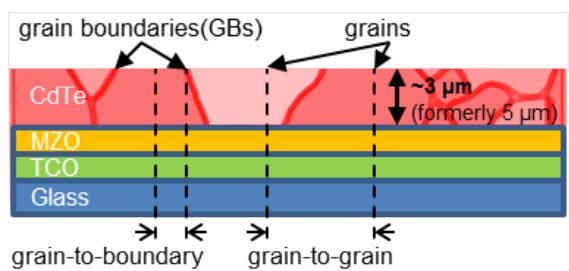

**Figure 1.** Schematic of CdTe samples examined in this study. Grains and grain boundaries are illustrated to define grain-to-grain and grain-to-boundary designations.

Figure 2 shows maps of the local *IE* of untreated and CdCl$_2$-vapor-treated CdTe (CdCl$_2$-CdTe), as measured with PEEM, before and after air exposure. Color scales in Figures 2(a-d) were adjusted independently to highlight contrast for each image. Grayscale insets in Figures 2(a-d) present the local *IE* according to a common scale shown on the left of Figure 2a. Details on how the local *IE* was determined from photoemission threshold measurements are provided in the Experimental section and in ref. 19.

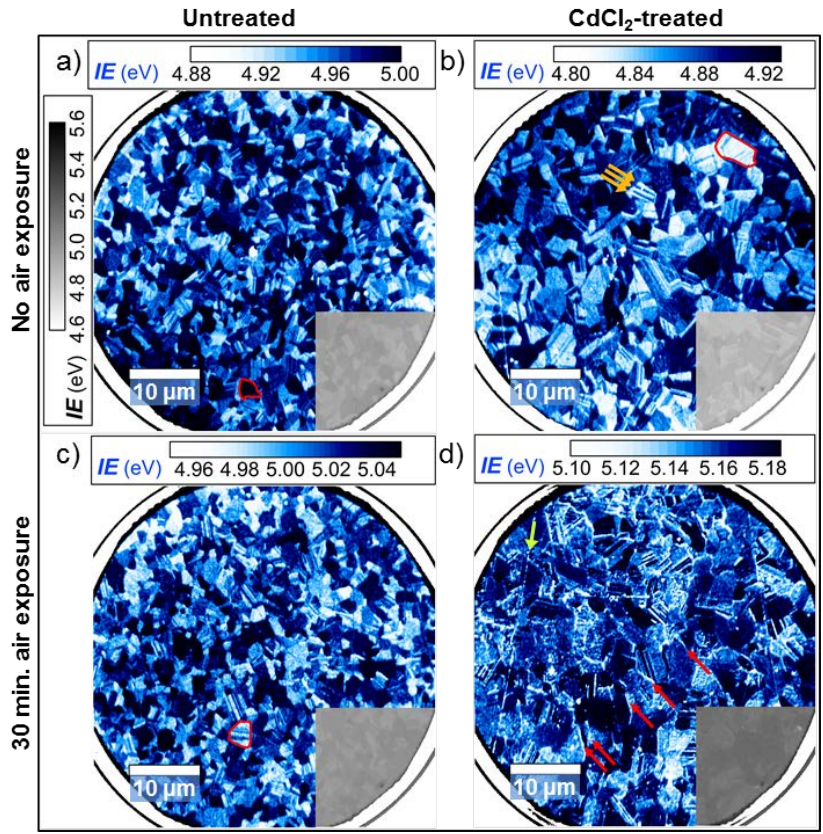

**Figure 2**. Maps of the ionization energy of (a,c) untreated and (b,d) CdCl$_2$-treated CdTe, before (top row) and after (bottom row) air exposure. Grayscale insets show the local ionization energy according to a common energy scale shown in (a).

The range of local *IE*s determined using PEEM was 4.8-5.2 eV and agrees with those measured using ultraviolet photoemission spectroscopy.[20,21] Micron-scale grain domains were evident in all *IE* maps (e.g. enclosed red lines in Figure 2). The grain sizes observed here (1-5 μm) are consistent with those observed in planarized polycrystalline CdTe films grown at higher temperatures.[8] All maps show parallel narrow domains (e.g. orange arrows in Figure 2b) associated with planar defects and twin boundaries.[8] Polish marks are distinguishable as long streaks (e.g., green arrow in Figure 2d).

All films showed grain-to-grain variation of *IE*. This variation can stem from: (i) a surface dipole effect arising from different CdTe facets exposed by the planarization of differently-oriented CdTe grains in the film,[22] or (ii) bulk effects, such as microscale band gap variations.[23] Both interpretations are consistent with reported variations apparent in grain-to-grain photocurrent[24] and Te concentration.[25]

As expected, $CdCl_2$ treatment tended to lower the overall *IE* of CdTe, but contrast enhancement at grain boundaries was surprisingly not observed as in other studies.[5,11,13,26] However, after air exposure of $CdCl_2$-CdTe (Figure 2d), the average *IE* increased and GB contrast in the *IE* map appeared. This contrast is visible as bright, venous lines decorating the edges of grains. For clarity, some of these GBs are pointed out by red arrows in Figure 2d. In contrast to previous microscopic studies of GBs in $CdCl_2$-CdTe, our results show that grain-to-boundary variation emerged only for $CdCl_2$-treated CdTe after air exposure.

This grain-to-boundary *IE* variation after air exposure of $CdCl_2$-CdTe challenges models that attribute modification of GB properties to an electrostatic rigid shift in the local electronic structure. These models consider that charged interface states deplete carriers in the vicinity of GBs, which results in all bands ($E_{vac}$, $E_{CB}$, and $E_{VB}$) bending together, leaving the GB *IE* unchanged.[5,10,11] However, local *IE* maps alone cannot refute the possibility that such a rigid shift in the bands is valid for other GBs, nor can it provide other details about local band bending (e.g. directionality, magnitude), which are necessary to verify GB models.

Additional insight is provided by comparing the $E_{vac}$ maps of $CdCl_2$-CdTe before and after air exposure (Figures 3a and 3c, respectively) with *IE* maps at the same locations (Figures 3b,d). Color scales in Figures 3(a-d) were adjusted independently to highlight contrast for each image, with the individual scales shown at the top of each image.

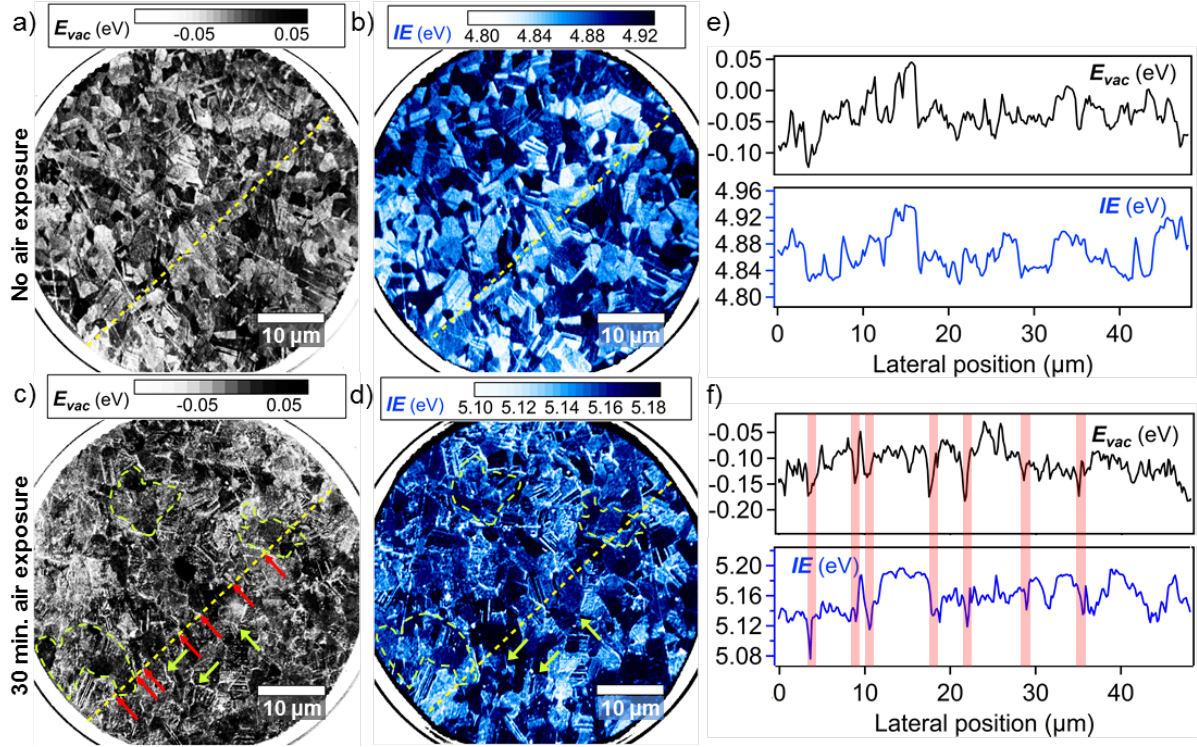

**Figure 3.** $E_{vac}$ (a,c) and $IE$ (b,d) maps of CdCl$_2$-treated CdTe, before and after air exposure. Each map has its own color scale, chosen to enhance contrast. (e) and (f) show the spatial variation of $E_{vac}$ and $IE$, before and after air exposure, as taken along the yellow dashed lines in the maps. For air-exposed CdCl$_2$-treated CdTe, regions of decreased $E_{vac}$ at GBs are denoted by red arrows in (c) and shaded in red in (f). GBs where a decreased $E_{vac}$ does not have a corresponding decrease in $IE$ are denoted by green arrows in (c) and (d). Green dashed lines in (c) and (d) illustrate the grains, which displayed no change in $E_{vac}$ or $IE$ at their boundaries relative to grain interiors.

Procedures to determine $E_{vac}$ from local photoemission spectra are provided in the Experimental section and in ref. 19. This side-by-side comparison clarifies the impacts of CdTe processing and air exposure on local electronic structure. To quantitatively compare the $E_{vac}$ and $IE$ variations, representative line profiles are plotted in Figures 3e and 3f. These profiles were taken along the yellow dashed lines shown in Figures 3(a-d).

For CdCl$_2$-CdTe exposed to air, red shaded areas in Figure 3f indicate regions of decreased $E_{vac}$ at GBs (highlighted by red arrows in Figure 3c).

For CdCl$_2$-CdTe with no air exposure (Figure 3e), electronic structure variations are predominantly step-wise in character, where $E_{vac}$ and *IE* largely change together spatially and energetically. The local $E_{vac}$ profile prior to air exposure features primarily grain-to-grain variation similar to the *IE* map. After air exposure (Figure 3f), $E_{vac}$ at GBs tended to decrease, but this decrease was not always associated with a commensurate decrease in *IE*. Assuming that the Fermi level was aligned throughout the sample, the differences in $E_{vac}$ represent differences in work function. Thus, air exposure produced lowered work functions at GBs as compared to their respective grain interiors.

A decreased work function at GBs in CdCl$_2$-treated CdTe[5,11,13,26] has been attributed to preferential segregation of chlorine (Cl)[15,25,27]. While it has been suggested that Cl dopes GBs with electrons and lower the work function relative to grain interiors (i.e. downward band bending),[5,11,26] the results presented here indicate to the contrary that incorporation of oxygen is also necessary for this GB "activation." The different incorporation of impurities at GBs as compared to grain interiors has been widely-observed for polycrystalline semiconductors. In particular, oxygen has been shown to incorporate differently at GBs in CuInSe$_2$.[28]

Some indications of an electrostatic rigid band shift are evident in Figure 3f, and appear as GBs with a downward shift in $E_{vac}$ and no changes in *IE* (e.g. green arrows in Figure 3c, d). However, other GBs show a concurrent decrease in both $E_{vac}$ and *IE*, while some GBs show no change at all in $E_{vac}$ or *IE* relative to grain interiors (e.g. regions in Figures 3c, d enclosed by green, dashed lines).

This suggests that additional grain boundary models are needed to account for all the observed electronic, chemical, and structural effects.[29] For example, previous studies observed that oxygen influences the electronic structure of GBs depending on their stoichiometry,[32-34] which may be Te- and/or Cd-depleted after CdCl$_2$ treatment.[11,15]

Focusing on the GBs with decreasing $E_{vac}$, an analysis of 172 representative GBs showed an average depletion width of 290 nm +/- 100 nm and an average energy barrier of 70 meV +/- 30 meV.[31,33] According to the Seto model,[34,35] these values equate to a net carrier density ($p_{net}$) of $10^{15}$ cm$^{-3}$, consistent with undoped CdTe, and a GB trap density ($p_{trap,GB}$) of $10^{11}$ cm$^{-2}$.[33] Estimates of $p_{trap,GB}$ are at or below the average trap density measured in CdCl$_2$-treated CdTe using admittance spectroscopy.[36] The agreements of $p_{net}$ and $p_{trap,GB}$ further establish the consistency of PEEM results with prior studies regarding undoped polycrystalline CdTe films after CdCl$_2$ treatment.

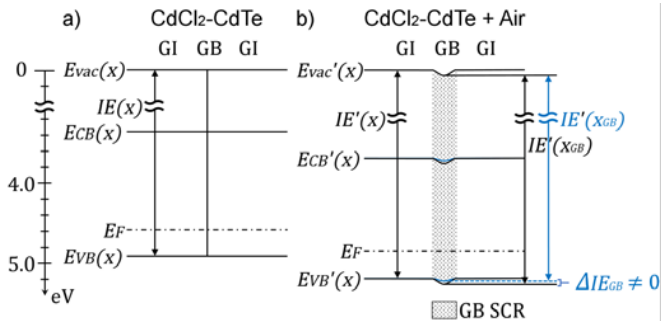

**Figure 4.** Schematic band diagram of grain interiors (GIs), grain boundaries (GBs), and GB space-charge regions (GB SCR) in CdCl$_2$-CdTe: (a) before and (b) after air exposure. Ionization energy, $IE(x)$, and vacuum level, $E_{vac}(x)$, are derived from PEEM measurements, which are referenced to a common vacuum level (i.e. $E_{vac}(x) = 0$ eV). The locations of the conduction band, $E_{CB}$, and the Fermi level, $E_F$, were determined from similar samples.[16] Colored (blue) labels in (b) indicate a change in GB ionization energy that appears in some locations after air exposure.

Figure 4 summarizes the effect of air exposure on grains and grain boundaries in the $CdCl_2$-CdTe surface, based on distributions of the local ionization energy and vacuum level at GBs measured using PEEM. Figure 4a represents the band structure in $CdTe$-$CdCl_2$ samples before air exposure. As noted previously, air exposure increased the average ionization energy *i.e.*, the location of the valence band ($E_{VB}$) shifts down the common energy scale after air exposure (Figure 4b). Furthermore, after $CdCl_2$ and air exposure "activate" GBs, the PEEM results indicate two possible scenarios regarding the space charge region near GBs. At some GBs, a decrease in the ionization energy occurs in tandem with downward band bending in $E_{vac}$, due to a different magnitude of bending in $E_{VB}$ and $E_{CB}$ than in $E_{vac}$. This scenario is denoted in Figure 4b by blue lines and labels. For other GBs, downward band bending in $E_{vac}$ occurs without a change in the ionization energy. This scenario is consistent with an electrostatic rigid shift in the local electronic structure, indicated by black lines and labels in Figure 4b.

CONCLUSION: In conclusion, PEEM measurements of $CdCl_2$-treated CdTe surfaces show that grain boundaries (GBs) display lower work function than the grain interior, confirming downward band bending near GBs, but only after $CdCl_2$-treated CdTe was exposed to air. This suggests that grain-to-boundary electronic structure variations commonly observed in CdTe involve interactions between multiple dopants and impurities. Additionally, the non-uniform activation of GBs in $CdCl_2$-treated CdTe after air exposure, as observed in vacuum level and ionization energy maps, indicates that not all GBs follow a doping-induced electrostatic rigid shift in the band structure suggested by prevailing GB models. Using PEEM as a direct probe of spatially-varying electronic structure, this study isolated the individual and cumulative effects of $CdCl_2$ processing and air exposure on local electronic structure in CdTe. Understanding these effects will enable the engineering of interface and bulk electronic structure to enhance photovoltaic performance.

METHODS: CdTe samples were fabricated on commercial TEC-10 superstrates consisting of soda lime glass coated with $SnO_2/SiO_2/SnO_2$:F, which were cleaned using a plasma process. $Mg_{0.23}Zn_{0.77}O$ (MZO) window layers were sputtered onto the superstrate, followed by closed-space sublimation of CdTe.[16] For $CdCl_2$-treated films, $CdCl_2$ vapor treatment[37] and thermal desorption of $CdCl_2$ salts were performed *in situ*.[38,39] During growth and $CdCl_2$ vapor treatment, substrate temperatures ranged between 425–500˚C. Source temperatures were 435–610˚C.

Sample surfaces were mechanically polished using a 1 μm diamond polish to remove topographical artifacts, resulting in root-mean-square and maximum surface roughness values of 4 nm and 39 nm, respectively, measured with atomic force microscopy. Sample surfaces were then cleaned using low-energy ion sputtering with 50 eV $Ar^+$ ions for 10–20 min. (~0.1-0.15 μA•$cm^{-2}$ fluence) to remove surface oxides and carbon.[17] Because of flat surface topography, artifacts caused by electric field distribution at sharp features were not considered in this study.[40] An inert environment sample transfer system was used to transfer samples between a dry $N_2$ (<0.1 ppm $O_2$, <0.1 ppm $H_2O$) glovebox and vacuum systems to avoid unintentional exposure of samples to air. The surface compositions of the films were verified after each step using x-ray photoelectron spectroscopy.[18] After all PEEM measurements, cross-sections of samples were examined with scanning electron microscopy to verify the polished CdTe layer thicknesses (3.2 μm and 2.5 μm for $CdCl_2$-treated and untreated films).

PEEM measurements were conducted in an Elmitec LEEM-III system with an electron energy analyzer and a tunable deep-ultraviolet (DUV) light source.[19] The spectral width of the DUV light was set to 50-100 meV throughout the measurement wavelengths ($h\nu$ = 3.6–7 eV). 600 × 600 sq. pixel PEEM images were acquired with a 48 μm field of view, for a pixel size of ~80 nm.

Using a fixed photon wavelength ($h\nu$ = 6.5 eV) local photoemission spectroscopy spectra were acquired by obtaining photoemission intensity images while sweeping the electron kinetic energy offset (10 meV steps) with respect to the energy window of the analyzer. The data collected are emission-angle integrated spectra for each pixel in the image. Fits to spectral edges specify locations of the $E_{vac}$ and $E_{VB}$.[19] $E_{vac}$ maps resulting from local fits were background subtracted to correct for the dispersion of the energy analyzer.

Local *IE* values were measured by recording the photoemission intensity at each pixel as a function of photon energy. At each pixel, *IE* was determined as the minimum photon energy above which photoemission intensity is observed (local photoemission threshold). The standard deviation of the signal acquired with lower energy photons was used as a criterion to define the local photoemission threshold.

Details regarding the analysis of ionization energies and grain boundary vacuum level profiles are provided in the Supporting Information.

## ASSOCIATED CONTENT

**Supporting Information**

The Supporting Information is available free of charge on the ACS Publications website at DOI: 10.1021/acsami.xxxxxxx.

Details about the extraction and analysis of grain boundary vacuum level profiles (PDF)

Comparison of ionization energy and vacuum level maps for treated and untreated CdTe films, before and after air exposure (PDF)

Histogram distributions of local ionization energy and vacuum level maps and untreated CdTe films, before and after air exposure (PDF)

## AUTHOR INFORMATION

**Corresponding Author**

*M. Berg. E-mail: mberg@sandia.gov.

*C. Chan. E-mail: cchan@sandia.gov.

**Author Contributions**

The manuscript was written through contributions of all authors. All authors have given approval to the final version of the manuscript. These authors contributed equally.

ACKNOWLEDGMENT

This work was supported by a U.S. Department of Energy, Office of Energy Efficiency and Renewable Energy SunShot Initiative BRIDGE award (DE-FOA-0000654 CPS25859), the Center for Integrated Nanotechnologies, an Office of Science User Facility (DE-AC04-94AL85000), a National Science Foundation PFI:AIR-RA:Advanced Thin-Film Photovoltaics for Sustainable Energy award (1538733), and Sandia National Laboratories Laboratory Directed Research and Development (LDRD). We thank R. Guild Copeland and Sergei A. Ivanov for providing technical assistance. Sandia National Laboratories is a multi-mission laboratory managed and operated by National Technology and Engineering Solutions of Sandia, LLC., a wholly owned subsidiary of Honeywell International, Inc., for the U.S. Department of Energy's National Nuclear Security Administration under contract DE-NA0003525.

TOC GRAPHIC:

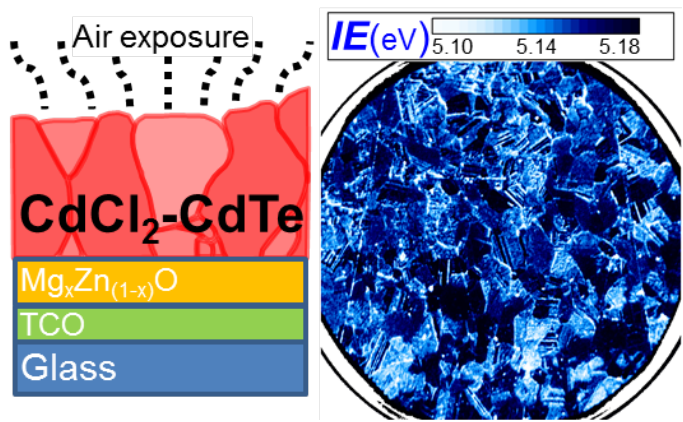

Supporting Information for

Local Electronic Structure Changes in

Polycrystalline CdTe with CdCl$_2$ Treatment and Air

Exposure

*Morgann Berg,[†,*] Jason M. Kephart,[‡,] Amit Munshi,[‡] Walajabad S. Sampath,[‡] Taisuke Ohta,[†]*

*Calvin Chan[†,*]*

[†] Sandia National Laboratories, Albuquerque, New Mexico 87185, United States

[‡] Department of Mechanical Engineering, Colorado State University, Fort Collins, Colorado

80523, United States

**Grain boundary profiles of CdCl$_2$-treated CdTe**

The electronic grain boundary model developed by Seto estimates the net doping density ($p_{net}$) and grain boundary (GB) trap density ($p_{trap,GB}$) given the depletion width ($w$) and the magnitude of band bending ($\Delta\Phi_{GB}$) at the space charge region (SCR) near GBs.[1] To perform this analysis, grain boundaries were identified in the vacuum level ($E_{vac}$) map of CdCl$_2$-treated CdTe after air exposure.

Grain-to-grain and grain-to-boundary contrast in the ionization energy (*IE*) and the corresponding maxima in the 1$^{st}$ spatial derivative of *IE* [$d(IE)/dx$] helped map the location of GBs. This GB map was then overlaid on the $E_{vac}$ map, and the SCRs near GBs were sampled by extracting line profiles oriented approximately orthogonal to grain boundaries, and spaced evenly from one another.

Figures S1a and S1b shows the perimeters of some grains (outlined in yellow) obtained from contrast in the *IE* map (Figure S1a) and the $d(IE)/dx$ map (Figure S1b) of CdCl$_2$-treated CdTe after air exposure. The GB map from each image match well. Figure S1c shows the complete GB map that resulted from this procedure, overlaid onto the $E_{vac}$ map of the same location. The locations of $E_{vac}$ line profiles taken along GBs, used to generate fits of the depletion width and barrier height of the space charge region (SCR) near GBs, are shown as blue lines in Figure S1c.

Of the 356 grain boundaries sampled, line profiles were either relatively flat, showed stepwise variation, or showed decrease in $E_{vac}$. Figure S2a shows examples of relatively flat, stepwise, and decreasing $E_{vac}$ at the grain boundaries. Line profiles that had a decrease (downward band bending) in the $E_{vac}$ at the location of the GB were fit with a Gaussian peak superimposed over a linear background to extract characteristics of the SCR. 172 of 356 grain boundaries exhibited this decrease in $E_{vac}$ and were fitted.

The FWHM of the Gaussian peak corresponded to the depletion width ($w$) of the SCRs near GBs. The amplitude of the Gaussian peak was associated with the magnitude of band bending at the GB ($\Delta\Phi_{GB}$).

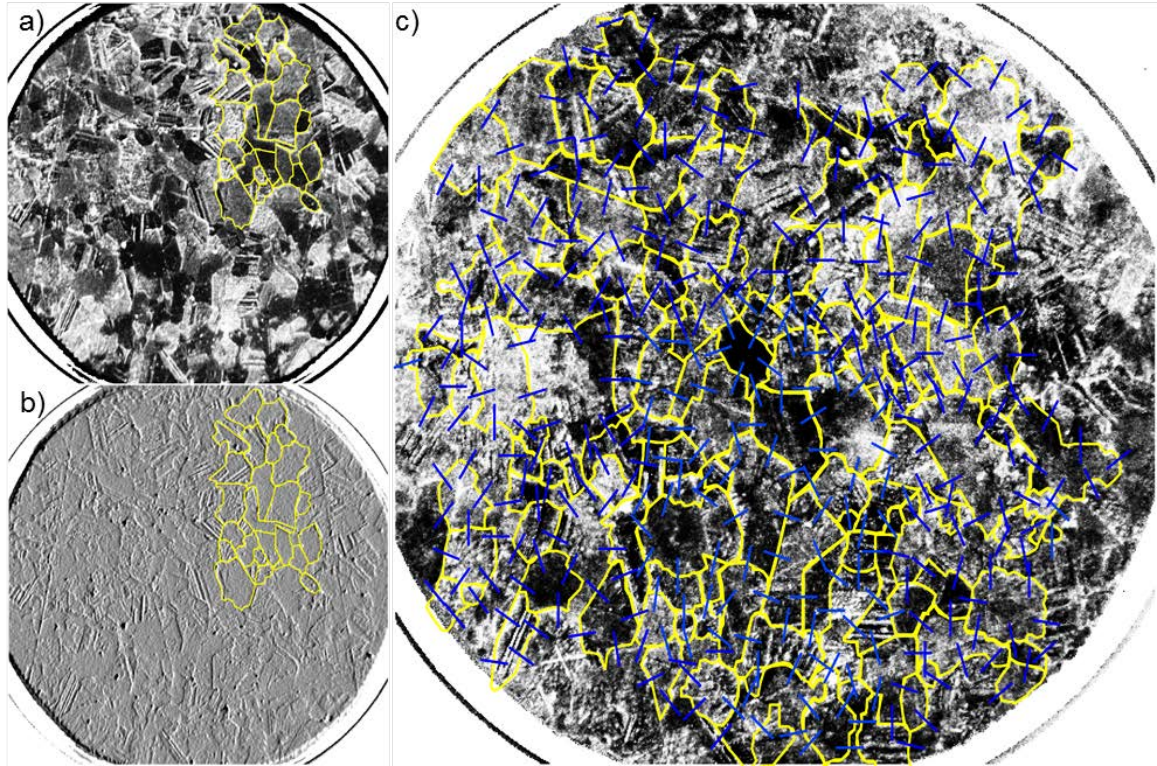

**Figure S1.** Maps of the (a) ionization energy ($IE$) and (b) $d(IE)/dx$ of CdCl$_2$-treated CdTe after air exposure, with some grain domains outlined in yellow. The derivative was taken along the horizontal $x$ direction. (c) Perimeters of identified grains are outlined by yellow lines overlaid on the corresponding vacuum level ($E_{vac}$) map. The location of line profiles sampled along grain boundaries (GBs) are shown in blue lines. $IE$ and $E_{vac}$ maps were obtained from local measurements of the photoemission threshold and photoemission spectra, respectively. The field of view for maps was 48 μm.

Figures S2b and S2c show histogram analyses of $w$ and $\Delta\Phi_{GB}$ obtained from the 172 analyzed GBs. The distribution of $\Delta\Phi_{GB}$ is asymmetric about the average value of 70 meV, and has a standard deviation of 30 meV; the distribution of $w$ is asymmetric about the average value of 290 nm, and has a standard deviation of 100 nm. According to the grain boundary model developed by Seto,[1,2] values of net doping density ($p_{net}$) and GB trap density ($p_{trap,GB}$) are obtained from $w$ and $\Delta\Phi_{GB}$ using the following expressions:

$$p_{net} = \frac{2\varepsilon\varepsilon_0 \Delta\Phi_{GB}}{e^2 w^2}, \qquad p_{trap,GB} = \frac{1}{e}\sqrt{8\varepsilon_0\varepsilon(p_{net})\Delta\Phi_{GB}},$$

where $\varepsilon = 10.3$ is the dielectric permeability CdTe,[3] $\varepsilon_0$ is the dielectric constant, and $e$ is the elementary charge. These expressions yield $p_{net} \approx 10^{15}$ cm$^{-3}$ and $(p_{trap})_{GB} \approx 10^{11}$ cm$^{-2}$ for CdTe.

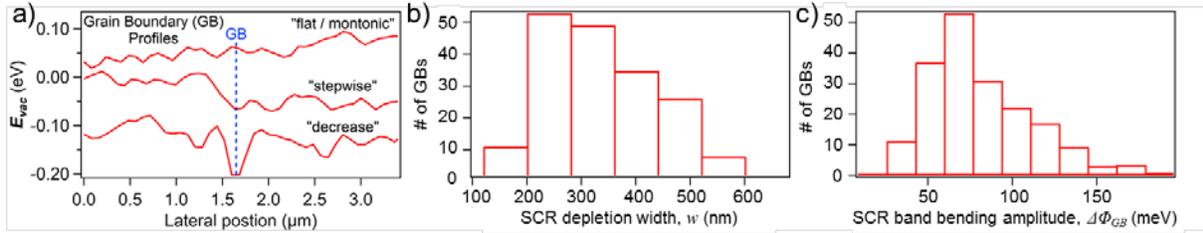

**Figure S2.** (a) Typical vacuum level variation observed at grain boundaries (GBs), and distributions of (b) the depletion width of the near-GB space charge region (SCR) and (c) the magnitude of band bending for "decreasing" (downward band bending) GBs.

**Vacuum level and ionization energy mapping of CdCl$_2$-treated and untreated CdTe, before and after air exposure**

As described in the experimental section and in refs. 4 and 5, local photoemission spectra and the photoemission yield curves provide maps of the relative vacuum level ($E_{vac}$) and ionization energy ($IE$). Variation of $E_{vac}$ indicates relative work function variation, assuming Fermi level alignment in the sample. $E_{vac}$ and $IE$ maps for CdCl$_2$-treated and untreated CdTe, before and after air exposure, are compared in Figure S3.

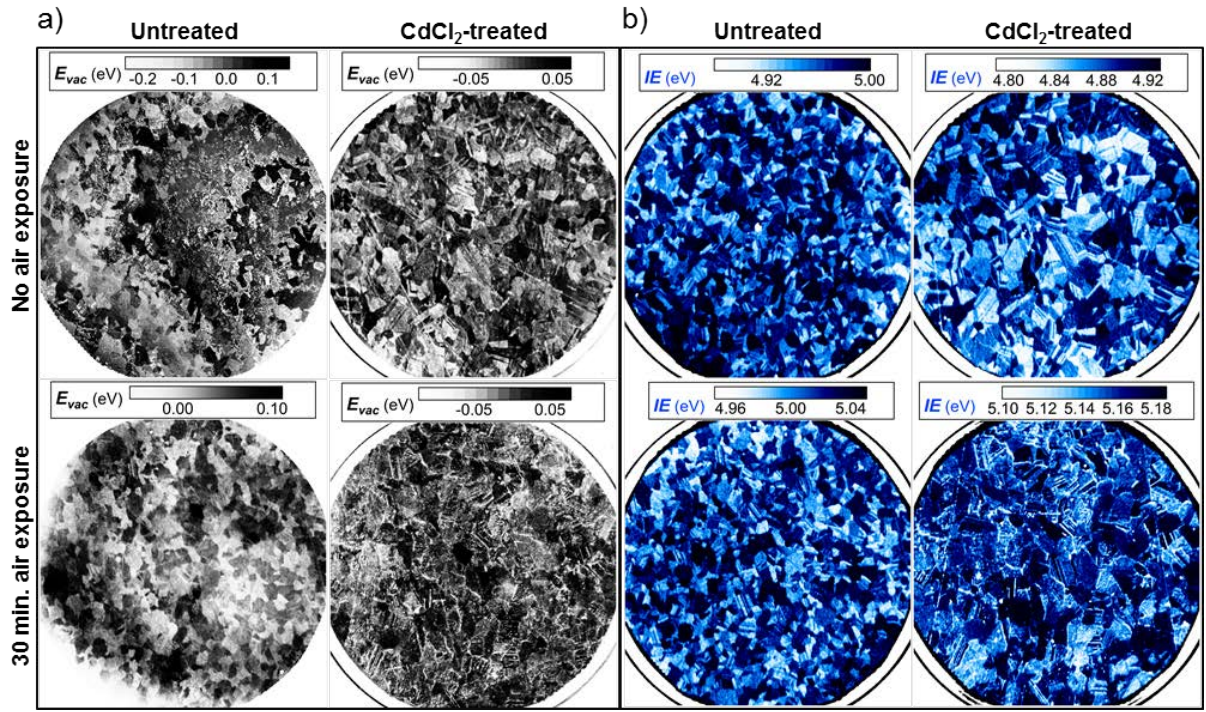

**Figure S3.** Maps of the (a) vacuum level and (b) ionization energy of untreated and CdCl$_2$-treated CdTe, before (top row) and after (bottom row) air exposure. Ionization energy ($IE$) and vacuum level ($E_{vac}$) maps were extrapolated from local measurements of the photoemission threshold and photoemission spectra, respectively. PEEM measurements of local photoemission threshold and photoemission spectra are described in ref. 5. Color scales for maps were chosen to highlight grain and grain boundary (GB) contrast. The field of view for maps was 48 μm.

$E_{vac}$ (Figure S3a) and *IE* (Figure S3b) maps of untreated CdTe films before and after air exposure show grain-to-grain variation similar to CdCl$_2$-treated CdTe prior to air exposure. Grain-to-boundary variation of the $E_{vac}$ and *IE* did not appear in CdTe until the CdCl$_2$-treated CdTe was exposed to air. $E_{vac}$ maps for CdCl$_2$-treated and untreated CdTe resembled photoemission intensity images previously reported in ref. 6. Aberrations from an unstable sample/holder contact affected the field distribution in the $E_{vac}$ map of untreated CdTe (Figure S3a, right column), similar to observations in ref. 6.

Histograms of these PEEM spectral maps help quantify and compare the extent of electronic structure variations. Figure S4 shows histograms for $E_{vac}$ (Figure S4a) and *IE* (Figure S4b) maps in Figure S3. The histograms were fitted using multiple Gaussian functions, denoted by dashed lines and numerical labels in Figure S4. Solid lines in Figure S4 indicate the overall fits.

Comparison of general trends (e.g. number of components, FWHMs) was the aim of this analysis, thus no specific physical parameters were imposed on the fittings. The general procedure for fitting was to include enough Gaussian fit functions to fit the edges of the histogram distribution, then introduce additional fit functions to improve the overall fit, if necessary.

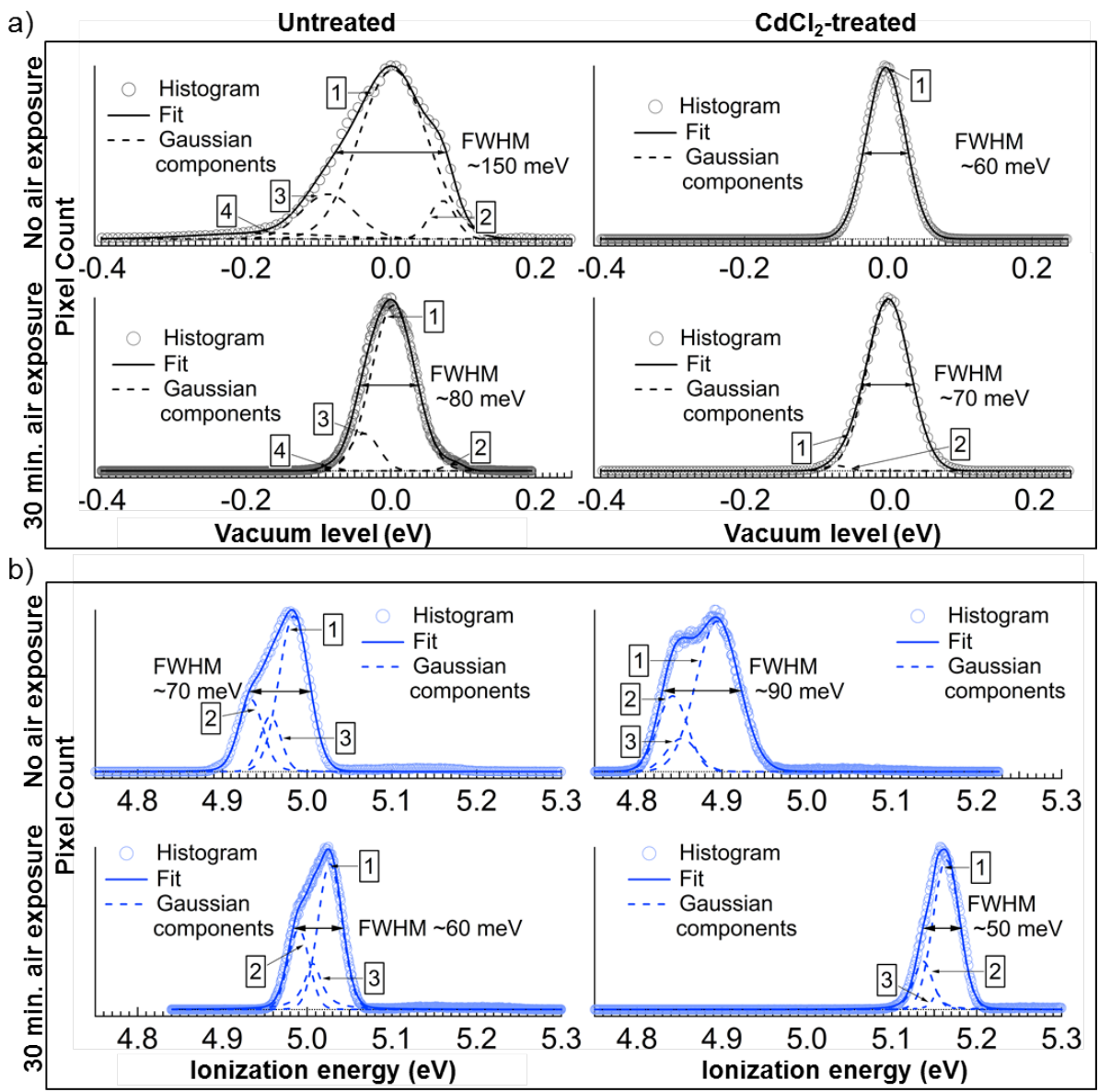

**Figure S4.** Histograms of (a) $E_{vac}$ and (b) $IE$ obtained from spectral maps of CdTe films with/without CdCl$_2$ treatment and with/without air exposure shown in Figure S4. Dashed lines with numerical labels indicate independent Gaussian fit functions, and solid lines correspond to the overall fits.

AUTHOR INFORMATION

**Corresponding Author**

*M. Berg. E-mail: mberg@sandia.gov.

*C. Chan. E-mail: cchan@sandia.gov.